\documentclass [prb,superscriptaddress, showpacs, twocolumn]{revtex4-1}
\usepackage{graphicx}
\usepackage{dcolumn}

\makeatletter

\newcommand{\Rmnum}[1]{\expandafter\@slowromancap\romannumeral #1@}
\makeatother

\begin{document}

\title{Change of magnetic ground state by light electron-doping in CeOs$_2A$l$_{10}$}

\author{D. D. Khalyavin}
\email{email: dmitry.khalyavin@stfc.ac.uk}
\affiliation{ISIS facility, STFC, Rutherford Appleton Laboratory, Chilton, Didcot, Oxfordshire, OX11-0QX,UK}
\author{D. T. Adroja}
\email{email: devashibhai.adroja@stfc.ac.uk}
\affiliation{ISIS facility, STFC, Rutherford Appleton Laboratory, Chilton, Didcot, Oxfordshire, OX11-0QX,UK}
\author{P. Manuel}
\affiliation{ISIS facility, STFC, Rutherford Appleton Laboratory, Chilton, Didcot, Oxfordshire, OX11-0QX,UK}
\author{J. Kawabata}
\affiliation{Department of Quantum Matter, ADSM, and IAMR, Hiroshima University, Higashi-Hiroshima 739-8530, Japan}
\author{K. Umeo}
\affiliation{Department of Quantum Matter, ADSM, and IAMR, Hiroshima University, Higashi-Hiroshima 739-8530, Japan}
\author{T. Takabatake}
\affiliation{Department of Quantum Matter, ADSM, and IAMR, Hiroshima University, Higashi-Hiroshima 739-8530, Japan}
\author{A. M. Strydom}
\affiliation{Physics Department, University of Johannesburg, P.O. Box 524, Auckland Park 2006, South Africa}
\date{\today}

\begin{abstract}
The effect of Ir substitution for Os in CeOs$_2$Al$_{10}$, with an unusually high Neel temperature of T$^*\sim$28.5K, has been studied by high-resolution neutron diffraction and magnetization measurements. A small amount of Ir ($\sim 8\%$) results in a pronounced change of the magnetic structure of the Ce-sublattice. The new magnetic ground state is controlled by the single ion anisotropy and implies antiferromagnetic arrangement of the Ce-moments along the $a$-axis, as expected from the anisotropy of the paramagnetic susceptibility. The value of the ordered moments, 0.92(1) $\mu_B$, is substantially bigger than in the undoped compound, whereas the transition temperature is reduced down to 21K. A comparison of the observed phenomena with the recently studied CeRu$_{1.9}$Rh$_{0.1}$Al$_{10}$ system, exhibiting similar behaviour [A. Kondo et al., J. Phys. Soc. Jpn. {\bf{82}}, 054709 (2013)], strongly suggests the electron doping as the main origin of the ground state changes. This provides a new way of exploring the anomalous magnetic properties of the Ce(Ru/Os)$_2$Al$_{10}$ compounds.
\end{abstract}


\maketitle

\indent The controversial interpretation of the phase transition recently found in the Kondo semiconductors CeRu$_2$Al$_{10}$\cite{ref:1} and CeOs$_2$Al$_{10}$\cite{ref:2,ref:2a} initiated their extensive study. After some debate,\cite{ref:2,ref:2a,ref:3,ref:4,ref:5,ref:6,ref:7} the nature of the transition was proven to be magnetic due to the ordering of the Ce-sublattice.\cite{ref:8,ref:9,ref:10} This ordering however is rather unusual and involves very small ordered moment along a direction not expected from the anisotropy of the static susceptibility ($\chi$), measured above the transition temperature. The anisotropy is very large and implies the $a$-crystallographic direction as the easy magnetic axis ($\chi_a>\chi_c>\chi_b$), whereas the neutron diffraction indicates the ordered moment 0.3 - 0.4 $\mu_B$ to be along the $c$-direction.\cite{ref:8,ref:11,ref:12} In addition, the ordering temperature is unexpectedly high T$^* \sim$28K, if one takes into account the moment value and the large distance $>5.2 \AA$ separating the Ce ions in the structure. Finally, a spin gap formation has been observed in inelastic neutron scattering experiments below the transition temperature in CeRu$_2$Al$_{10}$ and CeOs$_2$Al$_{10}$ as well as in the paramagnetic state of CeFe$_2$Al$_{10}$.\cite{ref:6,ref:10,ref:13,ref:13a}\\
\indent Aiming to elucidate the local electronic structure of Ce, Strigari et al.\cite{ref:14,ref:15} undertook polarization-dependent soft x-ray absorption measurements combined with magnetization data for both CeRu$_2$Al$_{10}$ and CeOs$_2$Al$_{10}$. The proposed crystal field ground state wave function provides quantitative agreement with the measured moment in high magnetic field applied along the $a$-direction and the small ordered moment along the $c$-direction found in the neutron diffraction experiments. The proposed ground state has been  confirmed directly from the inelastic neutron scattering measurements on CeOs$_2$Al$_{10}$, which reveals  two crystal field excitations at 40 and 60 meV.\cite{ref:15b} The mechanisms providing the a high ordering temperature and the gap formation are however unclear so far. Kimura et al.\cite{ref:15a} reported anisotropic changes in the electronic structure of CeOs$_2$Al$_{10}$ from polarized optical conductivity measurements and suggested that a charge density wave formation above T$^*$ is the primary instability which then induces anomalous magnetic ordering at T$^*$.\\
\indent Very recently, Kondo et al.\cite{ref:16} reported a remarkable change in the magnetic properties of CeRu$_2$Al$_{10}$ induced by a small amount ($\sim 5\%$) of Rh substitution onto Ru. In particular, a spin-flop like transition has been observed in a magnetic field $\sim$13T applied along the $a$-direction, indicating a new ground state with the ordered moments being along this axis as expected from the paramagnetic susceptibility. The observed effect has been attributed to the extra $4d$-electrons carried by Rh ions, suppressing the anisotropic character of the $c-f$ hybridisation and promoting localized state for the $4f$-electrons of Ce.\\
\indent The new ground state is of great interest and the primary questions immediately spring to mind: Does the doping result in any structural changes? Does it affect the magnetic propagation vector and the ordered moment size? In the present Rapid Communication, we address these matters for the CeOs$_{1.84}$Ir$_{0.16}$Al$_{10}$ system by means of high resolution neutron diffraction. We found that the small amount of Ir carrying extra $5d$-electrons indeed induces the new ground state with a considerably bigger ordered moment in comparison with the undoped counterpart. The moments are found to be along the $a$-axis, indicating predominance of the single ions anisotropy with the antiferromagnetic arrangement similar to that reported before for the CeOs$_2$Al$_{10}$ and CeRu$_2$Al$_{10}$ compounds.\\
\begin{figure}[t]
\includegraphics[scale=0.85]{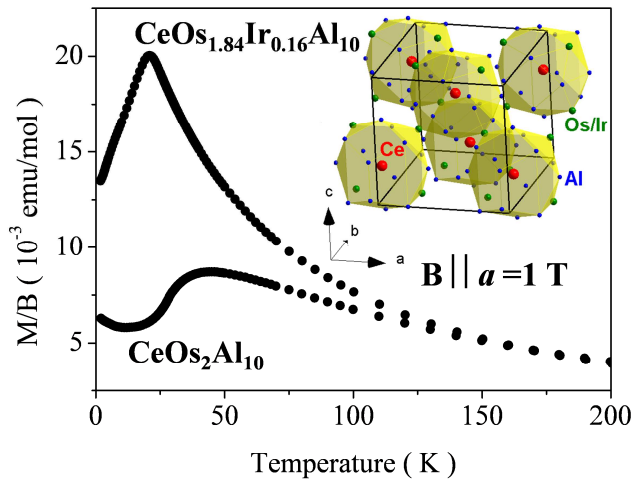}
\caption{(Color online) Magnetic susceptibility of Ce(Os$_{1-x}$Ir$_{x}$)$_2$Al$_{10}$ compositions, measured along the $a$-direction as a function of temperature. Inset shows schematic representation of the crystal structure of these compositions as a stacking of the CeAl$_{16}$(Os/Ir)$_4$ polyhedron cages.}
\label{fig:F1}
\end{figure}
\indent The polycrystalline sample of CeOs$_{1.84}$Ir$_{0.16}$Al$_{10}$ was prepared by ultrahigh-purity argon arc melting of stoichiometric quantities of the starting elements. The sample was annealed at 850C for 7 days for homogenization. A single crystal of the same composition was grown by the Al self-flux method.\cite{ref:2a} The neutron powder diffraction data were collected on the WISH time-of-flight diffractometer at the ISIS Facility of the Rutherford Appleton Laboratory, UK.\cite{ref:17} The sample (6g) was loaded into a cylindrical 6mm vanadium can and measured on warming between 1.5K and 30K in 3K steps using an Oxford Instrument cryostat. The crystal and magnetic structure Rietveld refinements were performed using FullProf program\cite{ref:18} against the data measured in detector banks at average $2\theta $ values of $58^o$, $90^o$, $122^o$, and $154^o$, each covering $32^o$ of the scattering plane. Magnetic susceptibility was measured using a SQUID magnetometer (Quantum Design MPMS) on both CeOs$_2$Al$_{10}$ and CeOs$_{1.84}$Ir$_{0.16}$Al$_{10}$ single crystals.\\
\indent The magnetic data obtained for the lightly Ir-doped crystal revealed a pronounced change in the low temperature susceptibility measured along the $a$-direction (Fig. \ref{fig:F1}). The changes are related to both the paramagnetic and the ordered phases indicating the behaviour similar to that recently observed by Kondo et al.\cite{ref:16} in the CeRu$_{1.9}$Rh$_{0.1}$Al$_{10}$ system. Above $\sim$150K, the paramagnetic susceptibilities for both doped and undoped crystals are very close. Below this temperature, the susceptibility for the undoped crystal starts deviating from the Curie-Weiss behaviour, whereas the Ir-doped crystal keeps it down to lower temperatures. Around T$^* \sim$21K, the susceptibility passes through a maximum indicating the phase transition. The critical temperature is essentially lower than in the undoped compound and at present it is unknown how it varies with the Ir content.\\
\begin{figure}[t]
\includegraphics[scale=0.72]{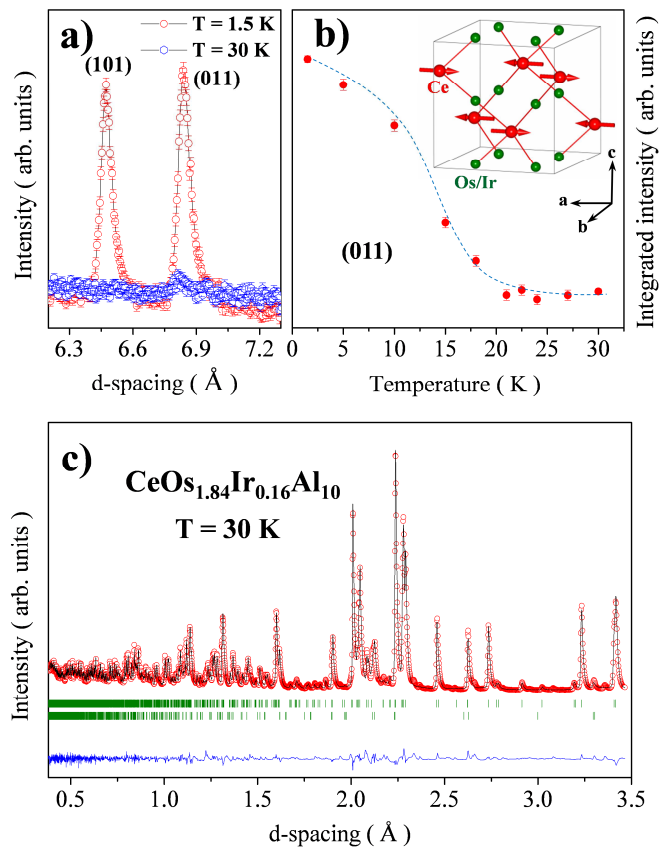}
\caption{(Color online) A part of the neutron diffraction patterns near the (101) and (011) magnetic reflections collected at different temperatures (a). Integrated intensity of the (011) magnetic peak as a function of temperature. The dotted line is a guide to the eyes (b). Inset shows the new magnetic ground state of CeOs$_{1.84}$Ir$_{0.16}$Al$_{10}$ (for clarity, only Ce and Os/Ir atoms are shown). Rietveld refinement of the neutron powder diffraction pattern collected at the backscattering detectors bank (average scattering angle is $154^o$) of the WISH diffractometer (c). The circle symbols (red) and solid line represent the experimental and calculated intensities, respectively, and the line below (blue) is the difference between them. Tick marks indicate the positions of Bragg peaks for the CeOs$_{1.84}$Ir$_{0.16}$Al$_{10}$ (top) and Os$_4$Al$_{13}$ (bottom) phases.} 
\label{fig:F2}
\end{figure}
\indent To directly explore the magnetic ground state of the electron doped CeOs$_{1.84}$Ir$_{0.16}$Al$_{10}$ system, high resolution neutron diffraction study has been performed. Although, the doping level is small, possible structural changes were explored in the refinement procedure. The parent Ir-free compound crystalizes into the orthorhombic YbFe$_2$Al$_{10}$ structure type with the $Cmcm$ space group.\cite{ref:19,ref:19a,ref:20} The structure encapsulates Ce into polyhedron cages formed by sixteen Al and  four Os atoms (Fig. \ref{fig:F1} inset), providing a large nearest-neighbour Ce-Ce distance. This structural model works well for all patterns collected at temperatures below and above the transition (Fig. \ref{fig:F2}c), so within the resolution limit of our diffraction experiment no structural modifications either compositional- or temperature-induced could be observed. A small amount of the Os$_4$Al$_{13}$ impurity phase has been identified and included into the refinement. The structural parameters obtained at T=30K are summarized in Table \ref{table:T1}. Evaluation of the unit cell parameters and the Ce-Ce bond distances also did not reveal any notable anomalies across the magnetic transition, indicating a weak magnetoelastic coupling in the system.\\
\begin{table}[t]
\caption{Structural parameters of CeOs$_{1.84}$Ir$_{0.16}$Al$_{10}$ refined from the neutron diffraction data collected at T=30K in the orthorhombic $Cmcm$ ($a$=9.1193(2)$\AA$, $b$=10.2554(2)$\AA$, $c$=9.1657(2)$\AA$, $R_{Bragg}$=4.18 $\%$) space group. Occupancies for all the atoms in the refinement procedure were fixed to the nominal chemical content.}
\centering 
\begin{tabular*}{0.48\textwidth}{@{\extracolsep{\fill}} c c c c c c} 
\hline\hline\\ 
Atom & Site & $x$ & $y$ & $z$ & $B_{iso}$ \\ [1.5ex] 
\hline\\ 
Ce & 4$c$ & 0 & 0.1260(6) & 0.25 & 0.23(8)\\ 
Os/Ir & 8$d$ & 0.25 & 0.25 & 0 & 0.40(4)\\
Al1 & 8$g$ & 0.2216(7) & 0.3661(7) & 0.25 & 0.2(1)\\
Al2 & 8$g$ & 0.3512(6) & 0.1305(7) & 0.25 & 0.4(1)\\
Al3 & 8$f$ & 0 & 0.1587(7) & 0.6005(6) & 0.2(1)\\
Al4 & 8$f$ & 0 & 0.3764(7) & 0.0485(6) & 0.2(1)\\
Al3 & 8$e$ & 0.2231(6) & 0 & 0 & 0.3(1)\\[1.5ex]
\hline
\hline  
\end{tabular*}
\label{table:T1} 
\end{table}
\begin{figure}[t]
\includegraphics[scale=0.98]{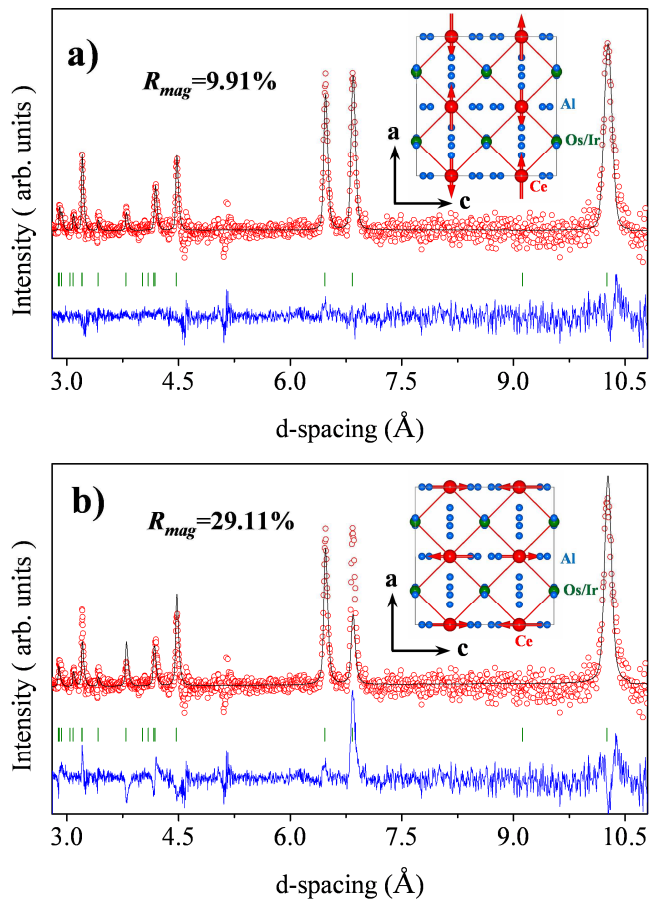}
\caption{(Color online) Rietveld refinements of the magnetic intensity of the CeOs$_{1.84}$Ir$_{0.16}$Al$_{10}$ composition obtained as a difference between the diffraction patterns collected at 1.5K and 30K. The circle symbols (red) and solid line represent the experimental and calculated intensities, respectively, and the line below (blue) is the difference between them. Tick marks indicate the positions of Bragg peaks for the magnetic scattering with the (${\bf k}=1,0,0$) propagation vector. The refinement quality is demonstrated for two models shown as insets, with moments along the $a$-axis (a) and $c$-axis (b).}
\label{fig:F3}
\end{figure}
\indent Magnetic Bragg scattering appears below T$^*\sim$21K resulting in a set of additional reflections clearly visible in a large d-spacing region (Fig. \ref{fig:F2}a). Their appearance correlates with a background reduction at low momentum transfer part of the patterns. This behaviour is expected in the case of a long-range magnetic ordering due to decreasing the paramagnetic scattering. All the magnetic peaks can be indexed with the ${\bf k}=(1,0,0)$ propagation vector, identical to the undoped CeOs$_2$Al$_{10}$ compound. The integrated intensities of these peaks as a function of temperature demonstrate a mean-field like behaviour consistent with the continuous nature of the transition. To refine the magnetic structure a difference between patterns collected at 1.5K and 30K was taken to better evaluate the fitting quality introduced by different models. The refinement procedure was conducted by the symmetry-based approach implying classification of the symmetrised pseudovector modes, localized on the Ce and Os/Ir positions, accordingly irreducible representations of the parent $Cmcm$ space group.\cite{ref:21,ref:22} Unique solution involving the single Y$^-_2$ representation (in the Miller and Love notations\cite{ref:22a}) and non-zero moments only for the Ce sites (Fig. \ref{fig:F2}b inset) has been identified based on the quality fitting of the magnetic intensities (Fig. \ref{fig:F3}a). The obtained model has the $C_Pmcm'$ magnetic symmetry ($\#$63.13.523 in the Litvin's classification scheme\cite{ref:23}) and is different from those reported for undoped CeOs$_2$Al$_{10}$ and CeRu$_2$Al$_{10}$ by the direction of the magnetic moments, which are along the $a$-axis in the new ground state. The type of the antiferromagnetic coupling violating the $C$-centering translation remains unchanged.  In particular, the ferromagnetic arrangement of Ce moment in the zigzag chain along the $c$ axis does not change between the pure and doped systems. This fact may suggest the importance of magnetic interaction along the zigzag chain via the hybridization with Os in the middle. The refined value of the Ce magnetic moments is substantially bigger 0.92(1) $\mu_B$ versus 0.29(1) in the undoped compound.\cite{ref:11} Refinement in the model with the moments aligned along the $c$-axis yields a much worse agreement with the experimental data (Fig. \ref{fig:F3}b) and can be unambiguously ruled out.\\
\indent Thus, the obtained result directly reveals the change of the ground state of the system upon Ir-substitution. The fact that this remarkable change has been observed in both Ce(RuRh)$_2$Al$_{10}$ and Ce(OsIr)$_2$Al$_{10}$ series provides a strong evidence that its main driving force is the electron doping rather that some crystal chemical effects. The value of the ordered moment 0.92(1) $\mu_B$ is in a good agreement with the high field magnetization measurements reported by Kondo et al.\cite{ref:16} in the undoped CeOs$_2$Al$_{10}$. The latter was found to be 0.95 $\mu_B$ along the easy $a$-axis. The  crystal field ground state wave function deduced by Strigari et al.\cite{ref:15} yields somewhat bigger moment along this axis $\sim 1.35 \mu_B$ indicating the presence of the Kondo screening effect. Finally, it should be pointed out that in spite of the bigger ordered moment and lower transition temperature in the new ground state of CeOs$_{1.84}$Ir$_{0.16}$Al$_{10}$, this composition still does not obeys the de Genne's scaling applied to the $Ln$(Os/Ru)$_2$Al$_{10}$ series ($Ln$-lanthanide), revealing anomalous mechanism providing the magnetic ordering in this class of the Ce-based materials.\\
\indent In conclusion, a small substitution of Os with Ir in CeOs$_2$Al$_{10}$ compound induces a pronounced change of the magnetic ground state. The new magnetic ordering involves the antiferromagnetic arrangement of Ce moments along the $a$-direction, as expected from the anisotropy of the paramagnetic susceptibility, with the ordered value being 0.92(1) $\mu_B$. The transition temperature is sensitive to the doping as well and is T$^*\sim$21K for the 8$\%$ of the Ir-content. The main origin of the observed doping effect is the extra $5d$-electrons carried by Ir. The obtained result demonstrates a great sensitivity of the Ce(Os/Ru)$_2$Al$_{10}$ systems to the carrier doping and provides a new way to study their anomalies magnetic properties.\\
\indent Acknowledgment: We would like to thank A.D. Hillier for interesting discussions. DTA would like to acknowledge financial assistance from CMPC-STFC Grant No. CMPC-09108. The work at Hiroshima University was supported by a Grant-in-Aid for Scientific Research on Innovative Area "Heavy Electrons" (20102004) of MEXT, Japan.


\thebibliography{}
\bibitem{ref:1} A. M. Strydom, Physica B {\bf{404}}, 2981 (2009).
\bibitem{ref:2} T. Nishioka, Y. Kawamura, T. Takesaka, R. Kobayashi, H. Kato, M. Matsumura, K. Kodama, K. Matsubayashi, and Y. Uwatoko, J. Phys. Soc. Jpn. {\bf{78}}, 123705 (2009).
\bibitem{ref:2a} Y. Muro, J. Kajino, K. Umeo, K. Nishimoto, R. Tamura, and T. Takabatake, Phys. Rev. B {\bf{81}}, 214401 (2010).
\bibitem{ref:3} M. Matsumura, Y. Kawamura, S. Edamoto, T. Takesaka, H. Kato, T. Nishioka, Y. Tokunaga, S. Kambe, and H. Yasuoka, J. Phys. Soc. Jpn. {\bf{78}}, 123713 (2009). 
\bibitem{ref:4} H. Tanida, D. Tanaka, M. Sera, C. Moriyoshi, Y. Kuroiwa, T. Takesaka, T. Nishioka, H. Kato, and M. Matsumura, J. Phys. Soc. Jpn. {\bf{79}}, 043708 (2010).
\bibitem{ref:5} K. Hanzawa, J. Phys. Soc. Jpn. {\bf{79}}, 043710 (2010). 
\bibitem{ref:6} J. Robert, J. Mignot, G. André, T. Nishioka, R. Kobayashi, M. Matsumura, H. Tanida, D. Tanaka, and M. Sera, Phys. Rev. B {\bf{82}}, 100404(R) (2010).
\bibitem{ref:7} K. Hanzawa, J. Phys. Soc. Jpn. {\bf{79}}, 084704 (2010).
\bibitem{ref:8} D. D. Khalyavin, A. D. Hillier, D. T. Adroja, A. M. Strydom, P. Manuel, L. C. Chapon, P. Peratheepan, K. Knight, P. Deen, C. Ritter, Y. Muro, and T. Takabatake, Phys. Rev. B {\bf{82}}, 100405(R) (2010).
\bibitem{ref:9} S. Kambe, H. Chudo, Y. Takunaga, T. Koyama, H. Sakai, T. U. Ito, K. Ninomiya, W. Higemoto, T. Takesaka, T. Nishioka, and Y. Miyake, J. Phys. Soc. Jpn. {\bf{79}}, 053708 (2010).
\bibitem{ref:10} D. T. Adroja, A. D. Hillier, P. P. Deen, A. M. Strydom, Y. Muro, J. Kajino, W. A. Kockelmann, T. Takabatake, V. K. Anand, J. R. Stewart, and J. Taylor, Phys. Rev. B {\bf{82}}, 104405 (2010).
\bibitem{ref:11} H. Kato, R. Kobayashi, T. Takesaka, T. Nishioka, M. Matsumura, K. Kaneko, and N. Metoki: J. Phys. Soc. Jpn. {\bf{80}}, 073701 (2011). 
\bibitem{ref:12} J. M. Mignot, J. Robert, G. Andre, A. M. Bataille, T. Nishioka, R. Kobayashi, M. Matsumura, H. Tanida, D. Tanaka, and M. Sera, J. Phys. Soc. Jpn. {\bf{80}}, SA022 (2011).
\bibitem{ref:13} J. Robert, J.-M. Mignot, S. Petit, P. Steffens, T. Nishioka, R. Kobayashi, M. Matsumura, H. Tanida, D. Tanaka, and M. Sera, Phys. Rev. Lett. {\bf{109}}, 267208 (2012).
\bibitem{ref:13a} D. T. Adroja, A. D. Hillier, Y. Muro, J. Kajino, T. Takabatake, P. Peratheepan, A. M. Strydom, P. P. Deen, F. Demmel, J. R. Stewart, J. W. Taylor, R. I. Smith, S. Ramos and M. A. Adams, Phys. Rev. B in press (2013).
\bibitem{ref:14} F. Strigari, T. Willers, Y. Muro, K. Yutani, T. Takabatake, Z. Hu, Y.-Y. Chin, S. Agrestini, H.-J. Lin, C. T. Chen, A. Tanaka, M. W. Haverkort, L. H. Tjeng, and A. Severing: Phys. Rev. B {\bf{86}}, 081105 (2012).
\bibitem{ref:15} F. Strigari, T. Willers, Y. Muro, K. Yutani, T. Takabatake, Z. Hu, S. Agrestini, C.-Y. Kuo, Y.-Y. Chin, H.-J. Lin, T. W. Pi, C. T. Chen, E. Weschke, E. Schierle, A. Tanaka, M. W. Haverkort, L. H. Tjeng, and A. Severing, Phys. Rev. B {\bf{87}}, 125119 (2013).
\bibitem{ref:15a} S. I. Kimura, T. Iizuka, H. Miyazaki, A. Irizawa, Y. Muro, and T. Takabatake: Phys. Rev. Lett. {\bf{106}} 056404 (2011).
\bibitem{ref:15b} D.T. Adroja et el to be published
\bibitem{ref:16} A. Kondo, K. Kindo, K. Kunimori, H. Nohara, H. Tanida, M. Sera, R. Kobayashi, T. Nishioka, and M. Matsumura, J. Phys. Soc. Jpn. {\bf{82}}, 054709 (2013).
\bibitem{ref:17} L. C. Chapon, P. Manuel, P. G. Radaelli, C. Benson, L. Perrott, S. Ansell, N. J. Rhodes, D. Raspino, D. Duxbury, E. Spill, and J. Norris, Neutron News {\bf{22}}, 22 (2011).
\bibitem{ref:18} J. Rodriguez Carvajal, Physica B {\bf{193}}, 55 (1993).
\bibitem{ref:19} V. M. T. Thiede, T. Ebel, and W. Jeitschko, J. Mater. Chem. {\bf{8}}, 125 (1998).
\bibitem{ref:19a} 2.A. I. Tursina, S. N. Nesterenko, E. V. Murashova, H. N. Chernyshev, and Y. D. Seropegin, Acta Crystallogr., Sect. E {\bf{61}}, i12 (2005).
\bibitem{ref:20} M. Sera, D. Tanaka, H. Tanida, C. Moriyoshi, M. Ogawa, Y. Kuroiwa, T. Nishioka, M. Matsumura, J. Kim, N. Tsuji, and M. Takata, J. Phys. Soc. Jpn. {\bf{82}}, 024603 (2013).
\bibitem{ref:21} H. T. Stokes, D. M. Hatch, and B. J. Campbell, ISOTROPY, stokes.byu.edu/isotropy.html (2007).
\bibitem{ref:22} B. J. Campbell, H. T. Stokes, D. E. Tanner, and D. M. Hatch, J. Appl. Crystallogr. {\bf{39}}, 607 (2006).
\bibitem{ref:22a} S. C. Miller and W. F. Love, Tables of Irreducible Representations of Space Groups and Co-Representations of Magnetic Space Groups, 4th ed. (Preutt Press, Boulder, 1967).
\bibitem{ref:23} D. B. Litvin, Acta Crystallogr., Sect. A: Found. Crystallogr. {\bf{64}}, 419 (2008).
\end{document}